# Modification of Contract Net Protocol(CNP) : A Rule-Updation Approach

Sandeep Kaur
Dept. of Comp. Sci. &Engg.,
BCET,Gurdaspur, PUNJAB,
INDIA

Harjot Kaur
Dept. of Comp. Sci. &Engg.,
GNDU Regional Campus,
Gurdaspur, PUNJAB, INDIA

Sumeet Kaur Sehra
Dept. of Comp. Sci. &Engg.,
GNDEC,Ludhiana, PUNJAB,
INDIA

*Abstract*—Coordination in multi-agent system is very essential, in order to perform complex tasks and lead MAS towards its goal.  Also, the member agents of multi-agent system should be autonomous as well as collaborative to accomplish the complex task for which multi-agent system is designed specifically. Contract-Net Protocol (CNP) is one of the coordination mechanisms which is used by multi-agent systems which prefer coordination through interaction protocols.  In order to overcome the limitations of conventional CNP, this paper proposes a modification in conventional CNP called updated-CNP. Updated-CNP is an effort towards updating of a CNP in terms of its limitations of modifiability and communication overhead.  The limitation of the modification of tasks, if the task requirements change at any instance, corresponding to tasks which are allocated to contractor agents by manager agents is possible in our updated-CNP version, which was not possible in the case of conventional-CNP, as it has to be restarted in the case of task modification. This in turn will be reducing the communication overhead of CNP, which is time taken by various agents using CNP to pass messages to each other. For the illustration of the updated CNP, we have used a sound predator-prey case study.

*Keywords—Multi-Agent System; Coordination; Communication Language; Norm-based Contract Net Protocol; utility parameters*

## I. INTRODUCTION

A multi-agent system (MAS) can defined to be a system comprised of multiple interacting intelligent agents [14] inside surroundings (environment). It can be accustomed to solve troublesome issues and problems that are not possible for a private agent or a monolithic system to unravel. So, it is having tremendous applications in any problem-solving domain as explained below. Intelligence is basically ability to reason, learn, act and react. Multi-agent system carries with and within it, its surroundings and agents. Generally multi-agent systems research refers to software system agents actively functioning, in order to achieve the goals of a multi-agent system or their individual goals. However, in a multi-agent system the agents could equally well be robots or human-beings. A multi-agent system may contain combined human-agent teams.

### A. Applications of MAS to real world

Multi-agent systems [13] are applied in the real world to graphical applications such as computer games. Agent systems have been used in films as well. They are also used for coordinated defence systems. Other applications include transportation, logistics, graphics, GIS , disaster management as well as in many other fields. It is widely being advocated for use in networking and mobile technologies, to achieve automatic and dynamic load balancing, high scalability, and self-healing networks.

In MAS, a single agent alone is not sufficient to solve any complex problem for which actually the MAS are designed, as it has not sufficient resources, information or competence. Therefore, in order to achieve the goals of the system and hence performing the tasks of the system, agent has to coordinate (cooperate) with rest of the agents of the system. And in order to ensure coordination (cooperation) [15], agents communicate with each in MAS by using various communicative acts of communication languages[1][2] like FIPA-ACL (Agent Communication Language), KQML (Knowledge Query and Manipulation Language), FLBC,UCL (Universal Communication Language), FACL (Form-based Agent Communication Language), and DBACL (Database Agent Communication Language). Also, this communication is governed by a set of protocols and coordination mechanisms. One of the best and oldest mechanisms used for coordination is CNP (i.e. Contract Net Protocol) [11][12].

There are numerous issues related to CNP, which have been modified and added later on to it. One is presented by Sun and WU in [6] , in which they have modified CNP by adding the concept of norms to conventional CNP and have termed it as Norm-based CNP by removing the limitation of conventional CNP of being inefficient to handle specialized interactions and hence, coordination.Another one is presented by Elmahalawyin [21], which is called Round Contract Net Protocol (RCont), which is modification of CNP using an acquaintance model. There are many versions of modifications available to CNP, and discussion of all them requires in itself a complete book. This paper is basically contributing to the modification of CNP by modifying one of its rules or phases, i.e.in the task processing phase, modification can be done at the level of manager agent for the task which is already being allocated to contractor agent corresponding to the changes in requirements of set of tasks which are to be performed by manager agents, which otherwise can only be performed by task repetition and restarting CNP from the beginning once again. This in turn will reduce the communicationoverhead or the time of processing required by CNP.

In addition to this, this modification is also studied comparatively in the agent communication languages, FIPA-ACL and KQML. To demonstrate the implementation of this modification; a predator prey case study is used. This paper is organized as follows. Section II is related to introduction of various coordination mechanisms used in MAS, Conventional Contract-Net Protocol (CNP), Norm-based Contract Protocol, as well as description of the limitations present in both of them. Section III is related to description of various agent communication languages, hence highlighting various importance differences between all of them. Section IV is related to the brief description of predator-prey case study and work done by us for modification of CNP by using proposed approach, and results derived out of this modification. Section V summarizes the conclusions.

## II. COORDINATION IN MAS

### A. Coordination

Coordination [15] comprises of a set of mechanisms necessary for the effective operation of Artificial Agent Societies (AASs). It is also defined as other process of managing dependencies between activities. Amongst its fundamental components, are the allocation of scarce resources; communication between the agents about intermediate results, coordination goals, capabilities and plans, status of the different aspects of the environment as well as providing some meta-level information. Coordination is required and is normally available also in cases in which there is not full cooperation amongst the agents or groups of agents. In a human society, for example, competition is constrained by consumer protection, various government agencies and antitrust laws. People and organizations antagonistic to one another may interact via prescribed legal channels. Coordination theory can be defined as a set of axioms, mathematical and logical constructs, and analytical techniques used to create a model of dependency management in AASs.

### B. Types of Coordination in MAS

According to Bergentti and Ricci [16], there are basically three main coordination approaches used in any MAS for managing coordination amongst a set of agents and they are based on the use of

- Tuple centres;
- Interaction Protocols; and
- Semantics of ACLs

In our case, for exercising coordination in MAS, agents are using Contract Net Protocol (CNP), which is one of the best approaches used for coordination, and is described in subsection below.

### C. Contract Net Protocol (CNP)

In multi-agent system, in order to accomplish any task agents need autonomy and collaboration (in terms of coordination). Contract net protocol (CNP) is coordination mechanism often used in a multi-agent system so as to coordinate amongst a set/group of agents. The original Contract Net Protocol (CNP), was originally developed by Smith and Davis [11].

CNP works like a business market where manager agent asks for bids from the contractor agents and then awards tasks to suitable contractor agent. In CNP, tasks are accomplished by breaking them down into sub-tasks by manager agent and then asking for bids for those sub-tasks from the contractor agents. Contractor agents replies with bids or refuse within a given deadline. Once deadline is reached, manager agent awards the task to the most suitable contractor agent having lowest bid.

Another minor modification of CNP is FIPA-Contract-Net-Protocol [5], in which there is addition of rejection and confirmation communicative acts. For the detailed study of CNP, the readers can refer[5].

Various limitations of Conventional CNP are in the form of attributes like responsiveness, load balancing, and fairness, utilization of resources, communication overhead, robustness, modifiability and scalability [17]. Although, all of them cannot be improved at the same time by enhancing CNP because this enhancement in itself will become a very complex problem. Here in our paper, we have tried to update CNP by modifying its rules used in communication to decrease the communication overhead, which will in turn increase the efficiency of CNP. This updated version of CNP is basically implemented using the predator-prey case study by modifying the rules of communication for predator agents.

## III. AGENT COMMUNICATION IN MAS

For achieving collaboration and hence coordination in MAS, agents interact with each other. And, for communication, they use various communicative acts of agent communication language (ACL), as; possibility for different agents to interact in an open environment heavily depends on the adoption of a common, standard Agent-Communication Language. The two most-widely used ACLs in practice are KQML and FIPA-ACL. But neither have yet been considered as standards as they are not capable of letting heterogeneous agents communicate as there are numerous other agent communication languages like Universal Communication Language(UCL), Database Agent Communication Language(DBACL), Formal Language For Business Communication(FLBC), Form –Based ACL(FACL) ) available for agent communication. These languages are actually application-specific languages, as their use varies from application-to-application. Like any other communication language, an Agent-Communication Language (ACL) also includes the definition of the syntax and the definition of the semantics.

Definition of the syntax is the way in which single words are put together and the definition of the semantics is the meaning of the communicative acts. By means of an agent-communication language, an agent can coordinate, communicate and exchange knowledge with other agents despite differences in their hardware platforms, operating systems, architectures, programming languages and representation and reasoning systems. Language is assumed to be the fundamental component of every interaction or communication. In a multi-agent environment, agents "talk" to each other by using an agent communication language.

For implementation of modified CNP, we have used a predator-prey case-study[3]. And, we have studied it comparatively, being implemented in FIPA-ACL and KQML languages. Therefore, this implementation and comparison will be incomplete without a brief introduction to these languages, which is described in subsections.

*A. FIPA-ACL*

The FIPA (Foundation for Intelligent Physical Agents) - Agent Communication Language (ACL) is based on speech act theory [13]: messages are actions, or communicative acts, as they are intended to perform some action by virtue of being sent. The specification consists of a set of message types and the description of their pragmatics (linguistics), i.e. the effects on the mental attitudes of the sender and receiver agents. Every communicative act is described with both a narrative form and a formal semantics based on modal logic. The specifications embrace steering to users who are already familiar with KQML in order to facilitate migration to the FIPA- ACL.The specification also provides the normative description of a set of high-level interaction protocols, including requesting an action, contract net protocol and several kinds of auctions [4][5].

*B. KQML*

KQML (Knowledge Query and Manipulation Language) is complementary to work on representation languages for domain content, including the DARPA Knowledge Sharing Initiative's Knowledge Interchange Format (KIF)[9]. KQML has also been used to transmit object-oriented data, and a wide range of information can be accumulated using it. KQML is a language for programs which use to communicate attitudes about information, such as querying, stating, believing, requiring, achieving, subscribing, and offering. KQML is indifferent to the format of the information itself, thus KQML expressions will often contain sub-expressions in other so-called content languages [10].

All the communicative acts, which are used by agents in MAS for interacting with each other, are governed by a set of rules and regulations, which are termed as agent communication protocols, as described in the next subsection below:

*C. Agent Communication Protocols*

Communication protocols [4] are widely recognized as a major and efficient concept to support many forms of interaction among agents, such as information sharing, task sharing, resource sharing, and coordination of actions, conflict resolution or commitments. When a set of agents interact through a protocol, each one is assumed to know when it may or must perform a communication act and what will be the effect of this performance[7][8]. Thus a protocol is a behavioural structure defined by:

- a set of (types of) communication acts feasible by agents;
- a set of roles that are played by agents;
- a set of behavioural rules stating under which circumstances an agent playing a particular role may or must perform a particular communication act. When an agent engages in a protocol, it chooses a role and commits to obey the protocol's rules.

Here, in this paper, for illustration of communicative acts and hence communication in FIPA-ACL and KQML, we are using once Contract Net Protocol (CNP) and then later on, its updated version in predator-prey case-study which is described in next section. Also, after this the comparative results of CNP and its updated version in both FIPA-ACL and KQML are analyzed graphically using various parameters used in CNP and these communication languages.

IV. WORK DONE

*A. A Predator-Prey Case Study*

We have used Predator-Prey case study for implementing the Contract-Net interaction protocol (FIPA-CNP) in FIPA-ACL and KQML languages, using their available per formatives (communicative acts), once using CNP and then its updated version, which is presented by us. The reason for using Predator-Prey system as a case study for studying the interaction in multi-agent systems is because it is very difficult to implement a real-world multi-agent system and study agent interaction in it. So, a directed test-bed or toy-domain like predator prey system is selected as a case study.

The Predator- Prey system is a pure-pursuit domain which involves multiple goal-oriented predator agents, prey agents and environment. The goal of predator agents is to chase and capture prey agents before it reaches the goal. The prey agent's goals are simply to evade the predator for a period of time, or to find and enter a goal square before they are captured. In the course of achieving goal, predator agents need to communicate with each other for passing information in space as they can communicate with each other about prey's location and form good strategy of capturing it. For the coordination between predator agents, we have used Contract-Net Protocol implemented once with conventional rules and secondly, using updated rules for decreasing the communication overhead in FIPA-ACL and KQML both. Both predators and prey cooperate to solve their "goals". The game takes place on an arbitrarily sized grid of squares [3].

Each predator or prey agent initially was completely autonomous. We have achieved this be writing and defining separate JAVA classes for both predator and prey agents. Later on, also they are also given ability to communicate through communicative acts of FIPA-ACL and KQML. Both of these are briefly described in the previous section. For implementation of these predator and prey agents, we have used JAVA-based platforms JADE (Java Agent Development Environment)[18][19] and JATlite [20].

*B. Description of Updated-CNP*

The conventional Contract-Net Protocol [11], coordinating the communication of contractor and manager agents, comprises of five phases as described below, In this we have also added our additional step to make it updated and efficient in its last and fifth phase which is also described after the description of CNP:

*1) Task Announcement:This phase is related to task announcement preparation by the manager agents for issuing them to every agent. This phase comprises of subtasks such as task abstraction, bid specification and expiration time specification. Task abstraction is the description of information related to tasks in abstract form which are to be*

*performed by contractor agents, for instance task name, task content description, bid specification is specifies the mandatory requirements which are to be fulfilled by contractor agents in order to be eligible to bid. Expiration time is the deadline for accepting the bid. This phase is also said to formulate in general, CFP(Call for Proposal).*

*2) Task Announcement Processing :* In this phase, according to the type of the task, the manager agent maintains a rank-ordered list of announcements that have been received and have not yet expired. Only those announcements which satisfy the bid specification criteria, will be one, to be allocated a rank and order in the list.

*3) Bidding :*This phase is related to contractor agents, in which they evaluate the tasks which are announced by manager agents and bid accordingly, if they meet all the necessary requirements specified in bid specification critieria.

*4) Bid Processing :*This phase is performed at the end of manager agent, who after receiving all the bids from the qualifying contractor agents, will evaluate them according to the task specification template, and then will process all the bids by ranking them. Then, from the ranked list or set, bids with the lowest cost are selected and tasks are allocated to them. This is further done, by informing to the contractor agents who had sent those bids, that their bids have been selected, and these particular set of agents will be now responsible for performing that particular set of tasks allocated to them. This all is performed by manager agents by making use of announced award message.

*5) Contract Processing, Reporting Results and Termination :* This is the last and the final phase of CNP, which actually marks the completion of CNP and its usage by contractor and manager agents for their task completion. In this phase, all the contractor agents who are allocated contracts, they are working for it, to complete the task allocated to them by means of the contract. During the processing of the contract, an information message is used for the general communication, whenever, it is occurring at any instance between the contractor and the manager agents. Also, in addition to that according to our updated-CNP, the task modification if any is required at the manager agent will be done be the same in the form of step 5 a) mentioned below, and for this again, manager agent will be informing contractor agent by making use of information message:

*a) Task Modification in CNP:*While the contract is processed by contractor agent in CNP, an interim report is sent by contractors to the manager agent, which will be summarizing work in progress or partially executed tasks which are being performed by the contractor agents. After, all the tasks are completed, a full report called final report is being submitted by contractor agents to manager agents, which will be containing a summary of all tasks which are allocated to contractors , there deadline, along with the time of completion.

But before, the final report is to be dispatched by contractor agent, the manager agents can send task change request to the contractors, if any change in task processing is required, i.e., any time before tasks are accomplished and results are sent back to the manager. Hence, this will be saving extra efforts incurred by task repetitions, if in case task change is done and again the commiuncation between the managers and the contractors is to be started from scratch. The changes in task can be anything from a contract setup between contractor and worker agents, or changes in bidding specification, as managers are only communicating with the eligible agents. Therefore, with this repetitions and hence wastage of efforts can be saved by changing requirement of tasks or terminating task execution at all. This will save communication overhead, which occurs in case of Conventional CNP if task repetition is done.

*C. Implementation with Case-Study*

We have implemented the predator-prey case-study, once with conventional CNP and then with updated CNP in both FIPA-ACL and KQML langauges. In case of old CNP it is not possible to change tasks once they are allocated to all predators to capture prey but in updated CNP we are asking some of the predator agents to change their goal and capture a specific prey.

In Predator prey system, an agreement of predators is intially done with environment where environment sends out prey's details while tasks of capturing prey are still in process. Environment can change the task of capturing prey to revoke chasing prey and work on something else like ignoring some of the preys (this is specifically in the case of stronger predator agents) and going after only a specific prey, which is more dangerous than others. In our implementation, it is the same set of predator agents, out of which few are working as manager predator agents and rest as contractor predator agents. All the feedback related to the prey agents in this case is provided by an environment. So, the initiation of CNP is between predator agents only. In case of old CNP it is not possible to change tasks once they are allocated to all predators to capture prey but in updated CNP, we are asking some of the predator agents to change their goal and capture a specific prey.

For the implementation of the above mentioned case-study and checking the performance of conventional and updated CNP in both FIPA-ACL and KQML languages, we have implemented predator and prey agents and hence system, once in JADE and then JATlite, these agents are communicating using communicative acts of once FIPA-ACL and then KQML. The coordination between predator agents for speeding up the process of prey-catching is done using CNP and updated-CNP. After the implementation, the performance is analyzed comparatively using graphs.

*D. Results*

The comparison of CNP and updated CNP by using three parameters, i.e., Updated Tasks, Tasks Repetitions and communication overhead in terms of time elapsed is illustrated in graphs, which are giving the summarized results of the execution of the predator-prey case study created with JADE and JATlite platforms for FIPA-ACL and KQML languages.The first set of graphs is related to performance of CNP and updated-CNP in case of predator-prey case study implemented in JADE, in which communication between predator and prey agents is done using FIPA-ACL performatives.

Graph1 in Figure 1 shows performance of CNP and updated-CNP for FIPA-ACL, while updation of tasks is done from the side of manager predator agents for contractor predator agents while CNP is being processed. Taken a set of 5 tasks, 2 were changed during execution.As, updated CNP accommodated almost all task changes immediately as compared to Conventional-CNP, so time was saved while processing of coordination between a set of predator agents, trying to chase and catch a prey.

Graph 2 in Figure 2 shows the performance of CNP and updated-CNP when task repetition was performed by contractor predator agents, the requirements corresponding to those tasks changed while CNP was in execution, so they were rescheduled by manager predator agents in conventional CNP, but in updated-CNP this change was absorbed within the protocol communication.

Graph 3 in Figure 3 shows the comparative performance of CNP and updated-CNP in terms of time taken for accomplishment of tasks after execution of tasks (5 tasks), including the changes made in tasks in between task execution.

The second set of graphs is related to performance of CNP and updated-CNP in case of predator-prey case study implemented in JATlite, in which communication between predator and prey agents is done using KQML performatives.

Graph 4 in Figure 4 shows performance of CNP and updated-CNP for KQML, while updation of tasks is done from the side of manager predator agents for contractor predator agents while CNP is being processed. Taken a set of 5 tasks, 2 were changed during execution. As, updated CNP accommodated almost all task changes immediately as compared to Conventional-CNP, so time was saved while processing of coordination between a set of predator agents, trying to chase and catch a prey.

Graph 5 in Figure 5 shows the performance of CNP and updated-CNP when task repetition was performed by contractor predator agents, the requirements corresponding to those tasks changed while CNP was in execution, so they were rescheduled by manager predator agents in conventional CNP, but in updated-CNP this change was absorbed within the protocol communication.

Graph 6 in Figure 6 shows the comparative performance of CNP and updated-CNP in terms of time taken for accomplishment of tasks after execution of tasks (5 tasks), including the changes made in tasks in between task execution.

## V. CONCLUSIONS

Coordination and Communication are two vital parts of MAS for its proper functioning, i.e. for performing a set of complex tasks and in order to fulfill its goals. For exercising coordination in MAS, it uses a set interaction protocols, CNP is one of them. The motive of CNP is to enhance communication between a set of agents (in the form of contractor and manager) which are using it. In case, if the task modification is required at the end of contractor agent for the task which is allocated by manager agent, then, it is only possible in conventional CNP after the termination of the protocol. Because, there is no means present in conventional CNP for task modification, only task repetition can be performed.

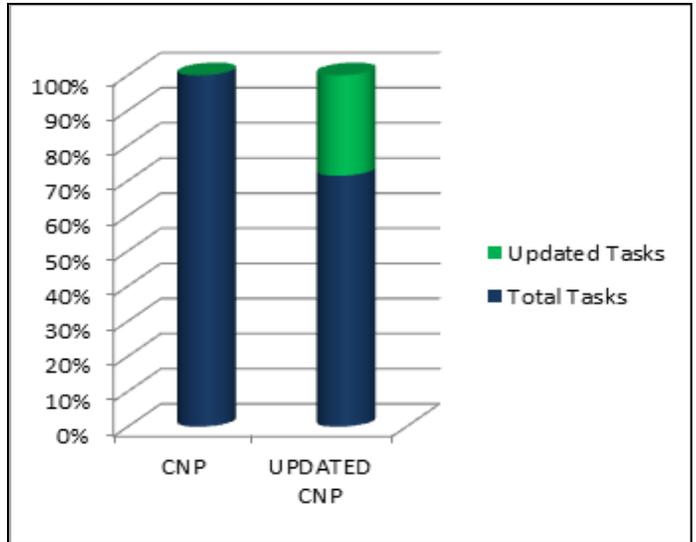

Fig. 1. Performance of CNP and updated CNP for task changes during CNP execution for FIPA-ACL.

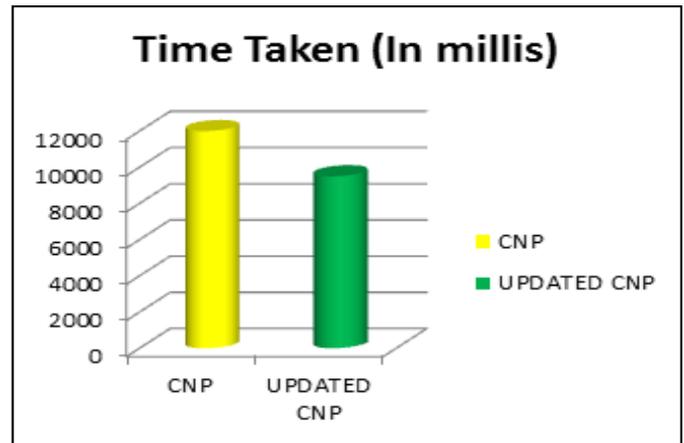

Fig. 2. Performance of CNP and updated CNP for task repetition during CNP execution for FIPA-ACL.

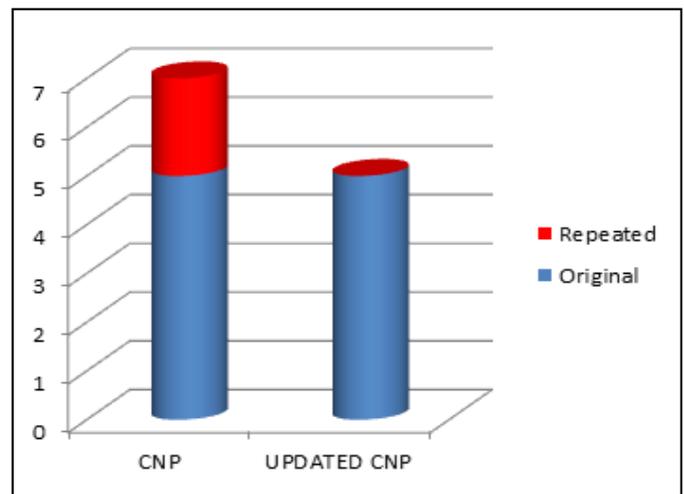

Fig. 3. Performance of CNP and updated CNP in terms of time for CNP execution for FIPA-ACL

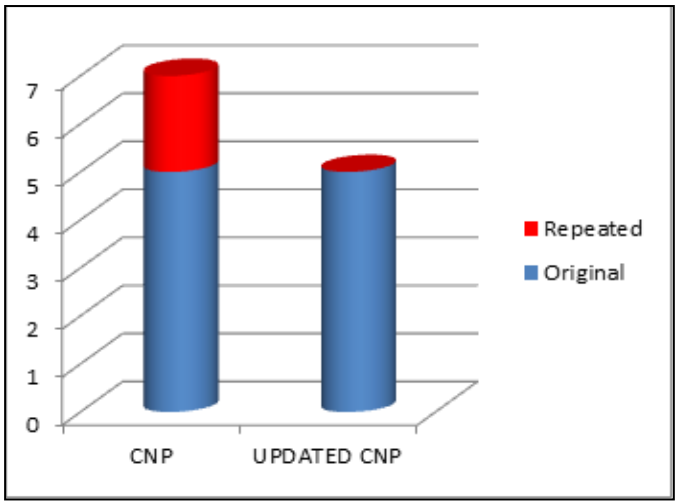

Fig. 4. Performance of CNP and updated CNP for task changes during CNP execution for KQML.

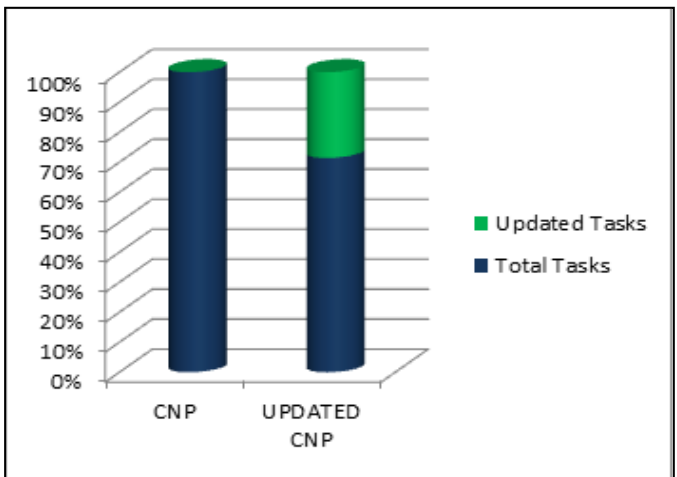

Fig. 5. Performance of CNP and updated CNP for task repetition during CNP execution for KQML

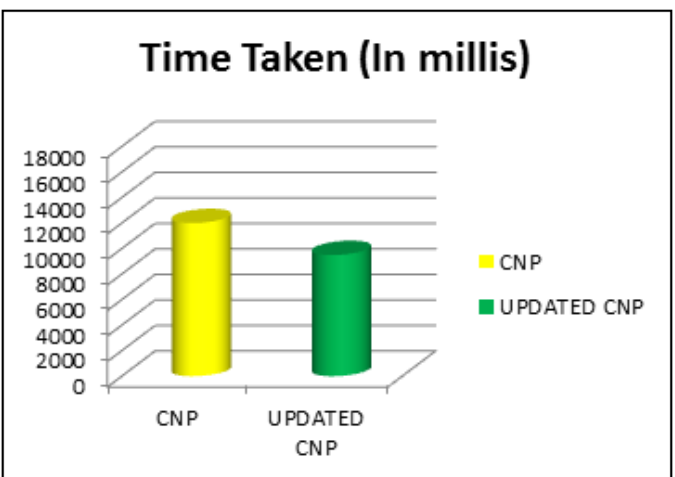

Fig. 6. Performance of CNP and updated CNP in terms of time for CNP execution for KQML

Here, in this paper, we have proposed and implemented an updated version of CNP, called updated-CNP. In updated-CNP, we have added an additional step into a conventional CNP, by the means of if any changes are to be made in task which is allocated to contractor agent by manager agent can be modified during the processing of the task before the final report of completion is send by contractor agent. This is save the overhead of restarting the process of CNP execution between a set of agents who wish to coordinate in order to achieve a certain objective. This will in turn improve the efficiency and effectiveness of protocol, in case there is frequent change in requirement set of manager agents, which, in turn requires task modification for contractor agents.

## VI. FUTURE WORK

In future, this work can be extended for other agent communication languages like FLBC, UCL and DBCL to know the performance of CNP and updated CNP, in case of application specific case studies. Then, comparison can also be performed for all these languages to check the relative performance of CNP and updated CNP in these languages.

In addition to this, scalability, load balancing, responsiveness, fairness, utilization of resources, robustnessand other limitations can be worked upon this using updated-CNP which we have demonstrated in this paper for different case studies depending upon the use of CNP in that case study. Also, all of these cannot be improved at the same time by enhancing CNP, because this enhancement in itself will become a very complex problem. Therefore, incremental enhancement of CNP can be done by removing one limitation at a time.